\documentclass[aps,prd,twocolumn,showpacs,floatfix,preprintnumbers,nofootinbib,superscriptaddress]{revtex4}
\usepackage{mathtools}
\usepackage{graphicx}
\usepackage{color}
\usepackage{natbib}
\usepackage{multirow}
\usepackage{amsmath}
\usepackage[amssymb]{SIunits}

\begin{document}

\title{A loophole to the universal photon spectrum in electromagnetic cascades:\\
application to the ``cosmological lithium problem'' }
\author{Vivian Poulin}
\author{Pasquale Dario Serpico}
\affiliation{LAPTh,  Universit\'e Savoie Mont Blanc, CNRS, B.P.110, Annecy-le-Vieux F-74941, France}

\date{\today}

\preprint{LAPTH-228/14}

\begin{abstract}
The standard theory of electromagnetic cascades onto a photon background
predicts a quasi-universal shape for the resulting non-thermal photon spectrum. This has been applied to
very disparate fields, including non-thermal big bang nucleosynthesis (BBN).  However, once the energy of the injected
photons falls below the pair-production threshold the spectral shape is much harder, a fact that has been
overlooked in past literature. This loophole may have important phenomenological consequences, since
it generically alters the BBN bounds on non-thermal relics: for instance it allows to re-open the possibility of purely electromagnetic solutions to the so-called
``cosmological lithium problem'', which were thought to be excluded by other cosmological constraints. We show this with a proof-of-principle example and a simple particle physics model, compared with previous literature.
\end{abstract}


\maketitle

\section{Introduction} 
Electromagnetic cascades, namely the evolution of $\gamma,\,e^\pm$ particle numbers  and energy
distribution following the injection of a energetic $\gamma$ or $e$ in a medium filled with radiation, magnetic fields and matter, is one of the physical
processes most frequently encountered in astroparticle physics, in domains as disparate as high-energy gamma-ray astrophysics, ultra-high-energy
cosmic ray propagation, or the physics of the early universe. In particular, the elementary theory of such a cascade onto a photon background
has been well known since decades, and can be shown via a textbook derivation (see Chapter VIII in~\cite{Ginzburg:1990sk}, for instance) 
to lead to a universal ``meta-stable'' spectrum---attained on timescales much shorter than the thermodynamical equilibration scale---of the form:  
\begin{equation} \label{eq:spectrum}
\frac{dN_\gamma }{dE_\gamma} = \left\{ \begin{array}{cl}
& K_0\left(\frac{E_\gamma}{\epsilon_X}\right)^{-3/2} \textrm{for } E_\gamma < \epsilon_X\,, \\
& K_0 \left(\frac{E_\gamma}{\epsilon_X}\right)^{-2} \:\:\;\textrm{for $\epsilon_X \leq E_\gamma \leq \epsilon_c$}\,, \\
& 0 \:\:\:\:\:\:\:\:\:\:\:\:\:\:\:\:\:\:\:\:\:\:\:\textrm{ for } E > \epsilon_c\,.
\end{array}\right.
\end{equation} 
In the above expression, $K_0 = E_0\epsilon_X^{-2}[2+\ln(\epsilon_c/\epsilon_X)]^{-1}$ is a normalization constant enforcing the condition that the total energy is equal to the injected electromagnetic energy, $E_0$;
 the characteristic energy $\epsilon_c= m_e^2/\epsilon_\gamma^{\rm max}$ denotes the effective
threshold for pair-production ($\epsilon_\gamma^{\rm max}$ being the highest energy of the photon background onto which
pairs can be effectively created);  $\epsilon_X \lesssim \epsilon_c/3 $ is the maximum energy of up-scattered inverse Compton (IC) photons. Natural units with $c=k_B=1$ are used throughout.

A notable application of this formalism concerns the possibility of a non-thermal nucleosynthesis phase in the early universe
(for recent review on this and other aspects of primordial nucleosynthesis, or BBN, see~\cite{Iocco:2008va,Pospelov:2010hj}).
The determination of the baryon energy density of the universe $\Omega_b$ inferred from the CMB acoustic peaks measurements can be used in fact to
turn the standard BBN into a parameter-free theory. The resulting predictions for the deuterium abundance (or  $^2$H, the most sensitive nuclide to $\Omega_b$) are in remarkable agreement with observations, providing a tight consistency check for the standard cosmological scenario. The $^4$He and $^3$He yields too are, broadly speaking, consistent with
this value, although affected by larger uncertainties. The $^7$Li prediction, however, is  a factor $\sim 3$ above its determination in the atmosphere of
metal-poor halo stars. If this is interpreted as reflecting a cosmological value---as opposed to a post-primordial astrophysical reprocessing, a question which is far from settled~\cite{Spite:2012us,Iocco:2012vg}---it requires a non-standard BBN mechanism, for which a number of possibilities have been explored~\cite{Iocco:2008va,Pospelov:2010hj}.  
 
In particular, cosmological solutions based on electromagnetic cascades have been proposed in the last decade, see for instance~\cite{Cyburt:2002uv}.
However,  typically they do not appear to be viable~\cite{Pospelov:2010hj}, as confirmed also in recent investigations (see for instance Fig. 4 in~\cite{Fradette:2014sza}, dealing with massive ``paraphotons'')
due to the fact that whenever the cascade is
efficient in destroying enough $^7$Li, the destruction of  $^2$H is too extreme, and spoils the agreement with the CMB observations mentioned above. Actually, this tension also affects some non-e.m. non-thermal BBN models, see for instance~\cite{Kusakabe:2014ola}.

This difficulty can be evaded if one exploits the property that $^7$Be (from which most of $^7$Li come from for the currently preferred value of $\Omega_b$, via late electron capture decays) has the lowest photodissociation
threshold among light nuclei, of about $1.59\,$MeV vs. $2.22\,$MeV for next to most fragile, $^2$H. Hence, to avoid any constraint from $^2$H while being still  able to photo-disintegrate some $^7$Be,  it is sufficient to inject photons with energy $1.6<E_\gamma/{\rm MeV}<2.2$, with a ``fine-tuned'' solution~(see e.g. the remark in~\cite{Pospelov:2010hj} or the discussion in~\cite{Kusakabe:2013sna}). Nonetheless,  it turns out to be hard or impossible to produce a sizable reduction of the final $^7$Li yield, while respecting other cosmological
bounds, such as those coming from extra relativistic degrees of freedom ($N_{\rm eff}$) or spectral distortions of the CMB. A recent concrete example of these
difficulties has been illustrated in~\cite{Ishida:2014wqa}, which tried such a fine-tuned solution by studying the effects of {\cal O(10)} MeV sterile neutrino decays. 

In this article, we point out that, depending on the epoch,  {\it at sufficiently low energies of injection the cascade develops differently and the final spectrum is significantly altered with respect to Eq.~(\ref{eq:spectrum})}, which has been incorrectly used till recently, see e.g.~\cite{Kusakabe:2013sna,Ishida:2014wqa}. As a concrete application, we show how this  re-opens a  window to a cosmological solution to the $^7$Li problem via e.m. decays. Additionally, one expects peculiar signatures associated
to such scenarios, which can be probed with cosmological observations. We will discuss this both in a proof-of-principle example and in the context of a
particle physics model, involving one sterile neutrino.  This was chosen for its simplicity and to allow for a direct comparison with the results of \cite{Ishida:2014wqa}, which studied a similar model.
Further considerations on some additional implications of our insight are finally outlined.

\section{ E.m. cascades and universal non-thermal spectrum} Our argument is the following: Let us assume that one injects  photons  at some time (or corresponding plasma temperature $T$) whose energy $E_0$ is below the pair production threshold at that epoch, which can be estimated for the CMB plasma to be  $\epsilon_c\sim m_e^2/(22\,T)\sim 10\, T_{\rm keV}^{-1}\,$MeV~\cite{Kawasaki:1994sc}. Note that {\it as long as $T < {\rm few}\,$keV, this is compatible with the typical nuclear photo-disintegration energies relevant for BBN}. It is clear that the spectrum of Eq.~(\ref{eq:spectrum}) cannot
stay valid in this regime: there is no pair-production cutoff, of course, but even the lower-energy part cannot be correctly captured by Eq.~(\ref{eq:spectrum}). Unless one considers {\it other physical processes} for the photon interactions, not included in the derivation of Eq.~(\ref{eq:spectrum}), there are no non-thermal electrons which can up-scatter CMB photons! 
Since the photon interaction probability is much smaller below pair production threshold, at leading level the injected spectrum below $\epsilon_c$ stays the same---apart for red-shifting, which happens on very long timescales with respect to particle photon interactions and hence we neglect. Accounting for the finite probability for the photons to scatter---via $\gamma\gamma$, via Compton scattering off the background electrons, 
or via Bethe-Heitler $e^\pm$ production onto background protons and Helium nuclei---one does end up with a suppression of the injected spectrum, plus 
a lower energy tail due to downgraded energy $\gamma$'s as well as $\gamma$'s produced via IC by the secondary $e$'s.  
The resulting secondary or tertiary photons, on the other hand, are typically at too low-energies to contribute to photo-dissociations and will
be neglected. Within this approximation, the Boltzmann equation describing the evolution of the distribution function $f_\gamma$ reads:
\begin{equation}\label{eq:BoltzmannPh}
\frac{\partial f_\gamma(E_\gamma)}{\partial t} = -\Gamma_\gamma(E_\gamma ,T(t))f_\gamma(E_\gamma,T(t))+{\cal S}(E_\gamma,t)\,,
\end{equation}
where ${\cal S}(E_\gamma,t)$ is the source injection term, $\Gamma$ is the total interaction rate, and we neglected the Hubble expansion rate~\footnote{In the case of the universal spectrum, this hypothesis of quasi-static equilibrium has been checked (see \cite{Kawasaki:1994sc}) and is in very good agreement with generated spectrum calculated numerically when this assumption is not made (see e.g. \cite{Svensson:1990} or \cite{Protheroe:1994dt}).}, since interaction rates are much faster and rapidly drive $f_\gamma$ to a quasi-static equilibrium, $\frac{\partial f_\gamma(\epsilon_\gamma)}{\partial t} =0$.
Thus, we simply have : 
\begin{equation}\label{eq:Spectre2}
f^{\textrm{S}}_\gamma(E_\gamma,t)= \frac{{\cal S}(E_\gamma,t)}{\Gamma_\gamma(E_\gamma)}\,,
\end{equation}
where the term ${\cal S}$  for an exponentially decaying species with lifetime $\tau_X$ and density $n_X(t)$, whose total e.m. energy injected per particle is $E_0$, can be written as 
\begin{equation}\label{eq:SSpectre}
{\cal S}(E_\gamma,t)=\frac{n_\gamma^0\zeta_X(1+z(t))^3\,e^{-t/\tau_X}}{E_0\tau_X}\,p_\gamma(E_\gamma)\,,
\end{equation}
with $z(t)$ being the redshift at time $t$, and the energy parameter $\zeta_X$ (conventionally used in the literature) is simply defined in terms of the initial comoving density of the $X$ particle $n_X^0$ and the actual one of the CMB, $n_\gamma^0$, via $n_X^0=n_\gamma^0\zeta_X/E_0$. A monochromatic emission line would then correspond to $p_\gamma(E_\gamma)=\delta(E_\gamma-E_0)$. For a two body decay $X\to \gamma\,U$ into a monochromatic line plus another not better specified (quasi)massless particle $U$, one would have $E_0=m_X/2$, where $m_X$ is the mass of the particle. Here, we will be interested in multi-MeV values for the mass $m_X$ and at temperatures of order few keV or lower, hence the thermal broadening is negligible and a Dirac delta spectrum as the one above is appropriate.

The interaction rate $\Gamma_\gamma$ is computed by accounting for: i)  Compton scattering over thermal electrons, $\gamma +  e_{th} \rightarrow \gamma + e$, taken from \cite{Kawasaki:1994sc}; ii) scattering off CMB photons: $ \gamma +  \gamma_{th} \rightarrow \gamma +\gamma$, for which we follow~\cite{Svensson:1990}; iii) Bethe-Heitler pair creation : $ \gamma +  N \rightarrow X + e^\pm $, for which we use the formulae of \cite{Jedamzik:2006xz}.
Note that we neglect the small effect due to the finite probability for the secondary or tertiary photons to induce some dissociations, i.e. once a photon interacts it is ``lost''. The results that we obtain are in this respect slightly conservative, by an amount which we estimated to be of the order of a few $\%$.

\section{ Non-thermal nucleosynthesis} 
\begin{figure}[!h]
\begin{center}
\includegraphics[width=0.4\textwidth]{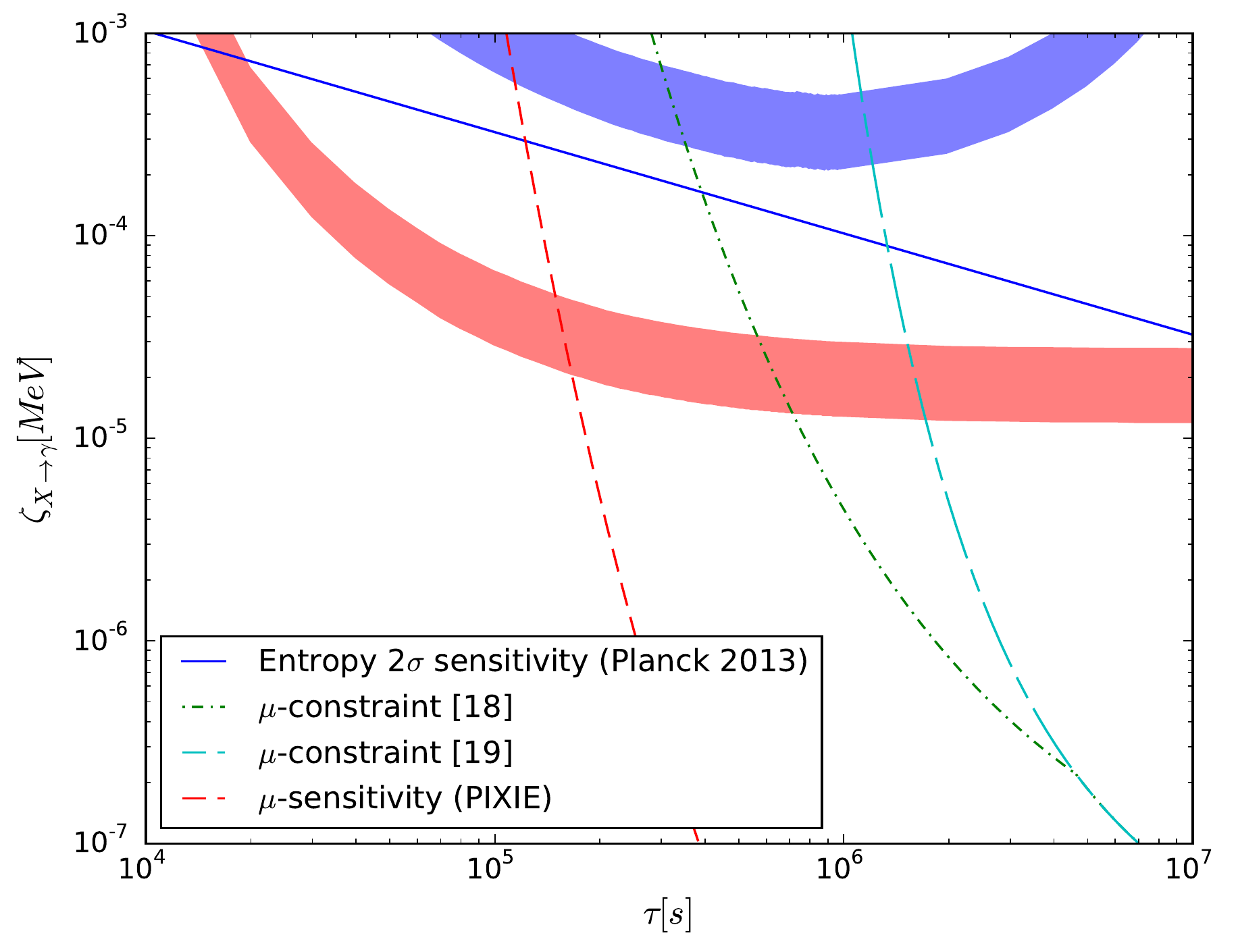}
\caption{{\footnotesize The lower band is the range of abundance parameter $\zeta_{X\rightarrow\gamma}$ vs. lifetime $\tau_X$, for which  the primordial lithium is depleted to 40\% to 70\% of its standard value,  for a monochromatic photon injection with energy $E_0=2\,{\rm MeV}$. The upper band  represents the analogous region  if we had distributed the same injected energy, up to $E_0=2\,$MeV, according to the erroneous  spectrum of Eq.~(\ref{eq:spectrum}). Above the solid blue curve, a change in entropy (and $\Omega_b$) between BBN and CMB time larger than the $2\sigma$ error inferred from CMB would be obtained. The region to the right of the dot-dashed green curve is excluded by current constraints from $\mu$-distortions in the CMB spectrum~\cite{Fixsen:1996nj} according to the computation of~\cite{Chluba:2011hw}, while the dashed cyan curve illustrates the weaker bounds that would follow from the less accurate parameterization of~\cite{Hu:1993gc}.
The dotted red curve is the forecasted sensitivity of the future experiment {\sf PIXIE}, corresponding to $|\mu| \sim 5\times10^{-8}$\cite{Kogut:2011xw}.\label{fig:ResultsBerylliumMonochromatic}}}
\end{center}
\end{figure}

At temperatures of few keV or lower, the standard BBN is over, and the additional nucleosynthesis can be simply dealt with  as a post-processing of the abundances computed in the standard scenario. 
The non-thermal nucleosynthesis due to electromagnetic cascades can be described by
a system of coupled differential equations of the type
\begin{eqnarray}
\frac{dY_A}{dt} & = & \sum_{{T}}Y_T\int_0^\infty dE_\gamma f_\gamma(E_\gamma) \sigma_{\gamma+T\rightarrow A}(E_\gamma) \nonumber \\
& - & Y_A  \sum_{{P}}\int_0^\infty dE_\gamma f_\gamma(E_\gamma) \sigma_{\gamma+A\rightarrow P}(E_\gamma)\label{nonthBBN}
\end{eqnarray} 
where: $Y_A \equiv n_A/n_b$ is the ratio of the number density of the nucleus $A$ to the total baryon number density $n_b$ (this 
factors out the trivial evolution due to the expansion of the universe); $\sigma_{\gamma+T\rightarrow A}$ is the photodissociation cross sections
onto the nuclei $T$ into the nucleus $A$, i.e. the production channel for $A$;  $\sigma_{\gamma+A\rightarrow P}$ is the analogous destruction channel (both cross sections
are actually vanishing below the corresponding thresholds). In general one also needs to follow secondary reactions of the nuclear byproducts of the photodissociation,
which can spallate on or fuse with background thermalized target nuclei (see for instance~\cite{Cyburt:2002uv}) but none of that is relevant for the problem at hand.
If the injected energy is $1.59<E_0/{\rm MeV}<2.22$, the only open non-thermal BBN channel is  $\gamma+{}^7\textrm{Be}\rightarrow {}^3\textrm{He} + {}^4\textrm{He}$, whose cross-section~\footnote{It is worth reporting that the cross-section for this process reported in
the appendix of \cite{Cyburt:2002uv} is erroneous. This has already been pointed out in~\cite{Ishida:2014wqa}, which we agree with. The correct formula
is used in the following.} we denote with $\sigma_\star$, there are no relevant source terms and only one evolving species (since $Y_7\ll Y_{3,4}$), thus yielding for the final (at $z_f$) to initial (at $z_i$) abundance ratio
\begin{equation}
\ln\left(\frac{Y_{{}^7\textrm{Be}}(z_i)}{Y_{{}^7\textrm{Be}}(z_f)}\right)  = \int^{z_i}_{z_{f}} \frac{n_{\gamma}^{0}\zeta_{X}\,\sigma_\star(E_{0})\,e^{\frac{-1}{2H_{r}^0\tau_X(z'+1)^{2}}} }{E_{0}H_{r}^0\tau_{X}\Gamma(E_{0},z)}
dz'\,.\label{result}
\end{equation}
To obtain Eq.~(\ref{result}), we transformed Eq.~(\ref{nonthBBN}) into redshift space, defining $H(z)=H_{r}^0(1+z)^2$ as appropriate for
a Universe dominated by radiation, with $H_{r}^0 \equiv H_{0}\sqrt{\Omega^{0}_{r}}$, 
  $H_0$ and $\Omega^{0}_{r}$ being the present Hubble expansion rate and fractional radiation energy density, respectively.
By construction, equating the suppression factor given by the RHS of the Eq.~(\ref{result}) to $\sim1/3$ provides a solution to the $^7$Li problem which is in agreement with all other constraints from BBN. In Fig.~\ref{fig:ResultsBerylliumMonochromatic}, the lower band shows for each $\tau_X$ the range of $\zeta_X$ corresponding to a depletion from 40\% to 70\%, for the case $E_0=2\,$MeV. Similar results would follow by varying $E_0$ by 10\% about this
value, i.e. provided one is not too close to the reaction threshold. The upper band  represents the analogous region  if we had distributed the same injected energy according to the  spectrum of Eq.~(\ref{eq:spectrum}), up to min[$\epsilon_c\,, E_0$]. It is clear that 
in the correct treatment a large portion of this region survives other cosmological constraints, described below, while none
survives in the incorrect treatment.

%%%%%%%%%            figure 1           %%%%%%%%%%%%%%
%%%%%%%%%%%%%%%%%%%%%%%%%%%%%%%%%
%%%%%%%%%%%%%%%%%%%%%%%%%%%%%%%%%
%%%%%%%%%%%%%%%%%%%%%%%%%%%%%%%%%
\section{CMB constraints} 
We mentioned the baryon abundances $\Omega_b$ inferred from CMB and BBN (notably $^2$H) probes are consistent within errors. This implies
that no major injection of entropy took place between the BBN time and the CMB epoch, otherwise the baryon-to-photon ratio (proportional to $\Omega_b$) would
have changed, see for instance~\cite{Feng:2003uy}. In a radiation-dominated Universe, the change in entropy associated to a release of energy into {\it all} e.m.
particles characterized by parameter $\zeta_{X\rightarrow e.m.}$ and a lifetime $\tau_X$ can be estimated as 
\begin{equation}
\frac{\Delta S}{S} \simeq \ln \frac{S_{f}}{S_{i}} = 2.14 \times 10^{-4}\frac{\zeta_{X\rightarrow e.m.}}{10^{-9}\textrm{ GeV}}\bigg(\frac{\tau_{X}}{10^6 \textrm{s}}\bigg)^{1/2}\,.
\end{equation}
For illustration, in Fig.~\ref{fig:ResultsBerylliumMonochromatic} the solid blue line represent the level of entropy release associated to a variation of $2\sigma$ around the best-fit measured value of $\Omega_{b}$ by {\sf Planck}, $\Delta S/S\simeq 0.022$~\cite{Ade:2013zuv}. Since the level of injected energy needed to solve the lithium problem via a monochromatic line is up to two orders of magnitude below the bounds, 
it is clear that this constraint is very weak, but for very short lifetimes of the order of $10^4\,$s.

Another constraint comes from the level of {\it spectral distortions} in the CMB. For the relatively short lifetimes relevant for the problem, the Compton scattering is fast enough that energy-redistribution is effective, no $y$-type distortion survives. On the other hand, processes that change the number of photons are relatively rare, and a residual distortion of the $\mu$-type is possible.  This has been constrained by  {\sf COBE-FIRAS} to be $|\mu| \leq 9\times10^{-5}$~\cite{Fixsen:1996nj}. 
 The level of spectral distortion produced by the decay process here has been estimated in the past (see for instance~\cite{Hu:1993gc}),
but a recent re-evaluation~\cite{Chluba:2011hw} found significant improvements at short lifetimes, essentially due to a better treatment of the time-dependence of the visibility function. The theoretical expectation for $\mu$ can be written as 
\begin{equation}
\mu \simeq 8.01 \times 10^2 \bigg(\frac{\tau_{X}}{1\textrm{ s}}\bigg)^{1/2}\,\bigg(\frac{\zeta_{X\rightarrow e.m.}}{1\textrm{ GeV}}\bigg)\, \mathcal{J}(\tau_X)\,,
\end{equation}
where the function $ \mathcal{J}$ is taken from~\cite{Chluba:2011hw}. The bound excludes the region to the right of the dot-dashed, green curve in Fig.~\ref{fig:ResultsBerylliumMonochromatic}. For comparison, the dashed cyan curve reports the much weaker bound that would follow from the approximations in~\cite{Hu:1993gc}.
We also checked that the extra constraint due to extra ``dark radiation'' parameterized by $N_{\rm eff}$ is irrelevant as long as the branching ratio in extra
relativistic species is not larger than {\it a couple of orders of magnitudes} with respect to the photon one. We thus conclude that there is a significant interval of lifetimes ($10^4\lesssim\tau_X/{\rm s}<10^6$) and corresponding energy injection parameter $10^{-3}>\zeta_{X\to \gamma}/{\rm MeV}>1.3\times 10^{-6}$ for which a perfectly viable solution is possible.  We remind once again that this possibility appeared to be closed due to the use of Eq.~(\ref{eq:spectrum}) beyond its regime of applicability. 
 
\begin{figure}[!t]
\centering
\includegraphics[width=0.4\textwidth]{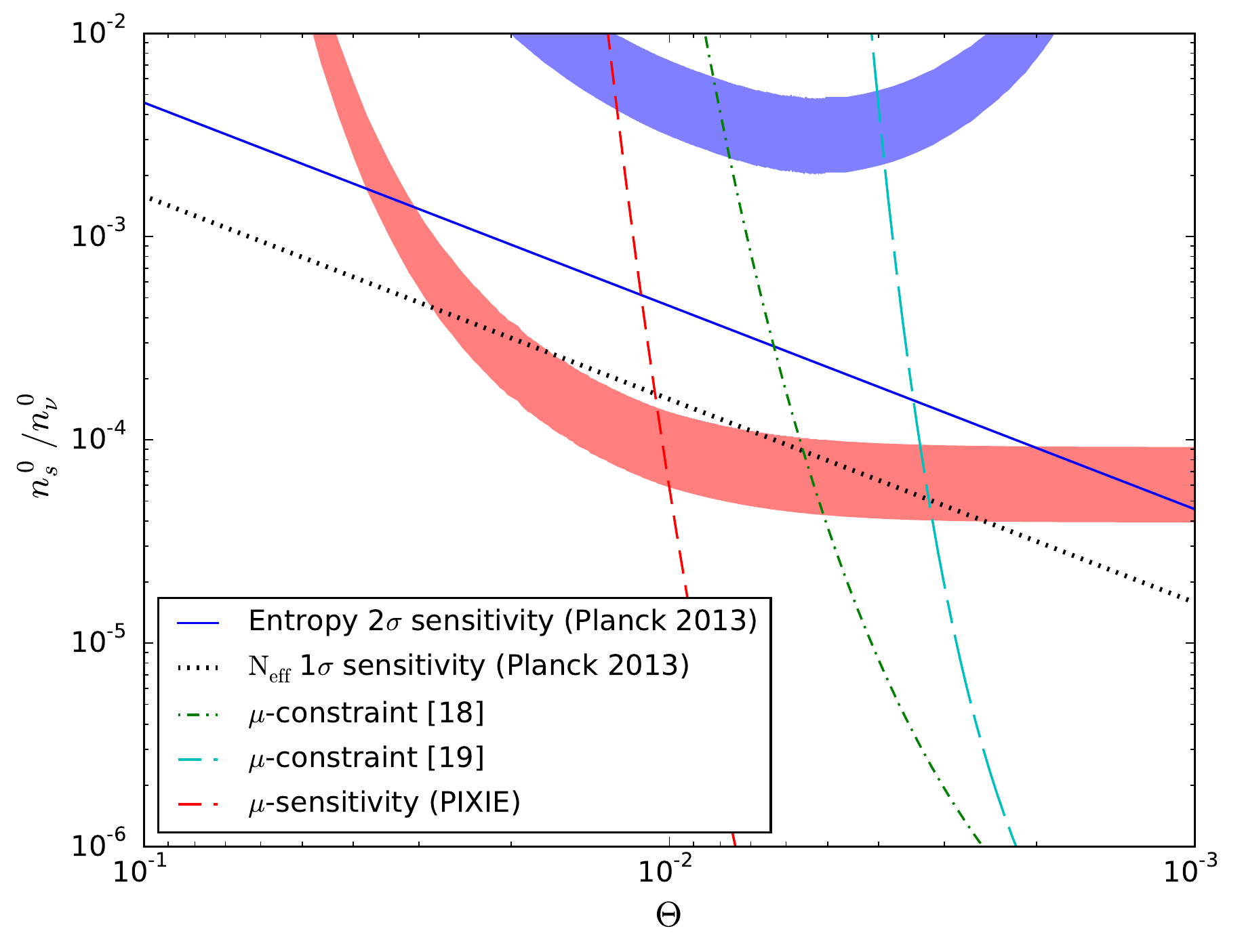}
\caption{\label{fig:ResultsSterileNeutrino} Constraints for the sterile neutrino model discussed in the text. The legend is the same as for case 1). }
\end{figure}

One may wonder how realistic such a situation is in a concrete particle physics model. Although we refrain here from detailed model-building considerations, it is worth
showing as a proof-of-principle that models realizing the mechanism described here while fulfilling the other cosmological constraints (as well as laboratory ones) can be actually constructed.
Let us take the simplest case of  a sterile Majorana neutrino with mass in the range $3.2<M_{s}/{\rm MeV}<4.4$, mixing with flavour $\alpha$ neutrinos via an angle $\theta_{\alpha}$. We also define  $\Theta^2 \equiv \sum_{\alpha}\theta_\alpha^2$. 
The three main decay channels of this neutrino are (see e.g.~\cite{Bezrukov:2009th} and refs. therein):
\begin{itemize}
\item $\nu_s\to 3\nu$, with rate $ \Gamma_{\nu_s\to 3\nu} \simeq \frac{G^2_F M_{s}^5\Theta^2}{192\pi^3}$;
\item $\nu_s\to \nu_\alpha e^+e^-$, with a rate depending on single $\theta_\alpha$'s;
\item $\nu_s\to \nu\gamma$, with a rate $\Gamma_{\nu_s\to \nu\gamma}\simeq \frac{9G^2_F\alpha M_{s}^5}{256\pi^4}\Theta^2$\,.
\end{itemize}
The resulting  branching ratios for the masses of interest and $\theta_e\ll \Theta$ are of the level of $0.9:0.1:0.01$, respectively.
It is physically more instructive to normalize the abundance of the $\nu_s$, $n_s^0$, in terms of one thermalized neutrino (plus antineutrino) flavour species, $n_\nu^0$. In Fig.~\ref{fig:ResultsSterileNeutrino}, we show the corresponding range of parameters in the $\Theta-n_s^0/n_\nu^0$ plane, for $M_s=4.4\,$MeV, for which the $^7$Li problem is solved, fulfills cosmological constraints and, provided that $\theta_e\ll \Theta$, also laboratory ones~\cite{Gelmini:2008fq}. It is worth noting that: i) the entropy release bound is now close to the region of interest, since the decay mode $\nu_s\to \nu_\alpha e^+e^-$, which is useless as far as the $^7$Be dissociation is concerned, dominates the e.m. energy injection. ii) A non-negligible fraction of relativistic ``dark radiation'' is now injected, mostly via the dominant decay mode $\nu_s\to 3\nu$;  hence we added the current $1\,\sigma$ sensitivity of {\sf Planck} to $N_{\rm eff}$~\cite{Ade:2013zuv}, with $\Delta N_{\rm eff}$ computed similarly to what done in~\cite{Ishida:2014wqa}. The needed abundance could be obtained in scenarios with low reheating temperature~\cite{Gelmini:2008fq}.

%%%%%%%%%            figure 2           %%%%%%%%%%%%%%
%%%%%%%%%%%%%%%%%%%%%%%%%%%%%%%%%
%%%%%%%%%%%%%%%%%%%%%%%%%%%%%%%%%
%%%%%%%%%%%%%%%%%%%%%%%%%%%%%%%%%

\section{Conclusions} We have discussed the breaking of the universality of the photon spectrum in electromagnetic cascades, when the energy of the injected
photons falls below the pair-production threshold. This may be of interest for a number of astroparticle applications, but in the specific case of the cosmological context, this happens when $E_\gamma\lesssim m_e^2/(22\,T)\sim 10\, T_{\rm keV}^{-1}\,$MeV. We noted that the energies concerned are of the same level of the binding energies of light nuclei. This implies a potential large impact on non-thermal nucleosynthesis models, notably of electromagnetic type, but could be also relevant for models with late time hadronic cascades. We provided an analytical estimate of the resulting (much harder) meta-stable spectrum of non-thermal photons, and 
showed that the impact is so large that it can potentially re-open the possibility of electromagnetic cacade solutions to the so-called
``lithium problem'', which were thought to be excluded by other cosmological constraints. We substantiated this point with a proof-of-principle example
of a photon line injection at $\sim 2\,$MeV from a particle decay, satisfying by construction all other BBN constraints but, not trivially, also all other
cosmological bounds plaguing previous attempts. Although we did not indulge into particle model building, we proved that the right conditions
can be actually  satisfied in a simple scenario involving a $\sim 4\,$MeV sterile neutrino mostly mixed with $\nu_\tau$ and/or $\nu_\mu$ with effective mixing angle  $\Theta\sim 10^{-2}$.

The possibility to find new mechanisms to deplete the standard BBN prediction of lithium abundance in a consistent way is probably the most spectacular consequence of our investigation.
In turn, this could stimulate more specific model-building activities.  For instance, decays of relatively light new neutral fermionic particles $X$ for which the $\nu+\gamma$ channel is the only two body standard model channel opened---as it is the case for the light gravitinos in supergravity models---constitute a natural class of candidates. Alternatively, one may think of decaying scenarios involving a pair of quasi degenerate mass states $X$ and $Y$, which are potentially much heavier than the MeV scale.
 Some of these scenarios may be motivated by other astroparticle or particle physics reasons and certainly deserve further investigation.
We also showed how improvements in the determination of $\mu-$type spectral distortions bounds of the CMB  might be crucial to test these scenarios: testing frameworks for the particle physics solutions to the lithium problem may thus provide additional scientific motivations for future instruments like {\sf PIXIE}~\cite{Kogut:2011xw}. Computations of distortions corresponding to specific injection histories may also be refined: for instance, for short lifetimes relativistic corrections to the double Compton and Compton scattering may be important to improve the theoretical accuracy~\cite{Chluba:2013kua}.

Finally, from a phenomenological perspective, an obvious spin-off of our work would be to re-compute the BBN bounds to electromagnetic decaying particles in cases where the universality of the spectrum of Eq.~(\ref{eq:spectrum}) breaks down. Preliminary results indicate that bounds can be easily modified by one order of magnitude. These results will be reported in  a forthcoming publication.

\acknowledgments
We thank J. Chluba and J. Pradler for comments on the manuscript. Support by the Labex grant ENIGMASS is acknowledged.

%%%%%%%%%%%%%%%%%%%%%%%%%%%%%%%%%%%%%%%%%%%%%%%%%%%%%%%%%%%%%%%%%%%%%%%

%%%%%%%%%%%%%%%%%%%%%%%%%%%%%%%%%%%%%%%%%%%%%%%%%%%%%%%%%%%%%%%%%%%%%%%
\end{document}